\documentclass{article}

\begin{document}
\baselineskip=1.5\baselineskip

\begin{center}
{\Large {\bf Scenarios for GCRT J1745-3009}}

S.B. Popov

Sternberg Astronomical Institute, Moscow, Russia
\end{center}

\begin{abstract}
 I discuss several scenarios to explain properties of the radio transient
source GCRT J1745-3009. Namely, a highly magnetized neutron star
 on the propeller or georotator
stage, a transient propeller, and an ejector in a binary system are
discussed. Simple populational estimates favor the transient propeller
model.
\end{abstract}

\section{Introduction}

Enigmatic radio source GCRT J1745-3009 was discovered in a 
dedicated search for transients in the galactic center region
few years ago (Hyman et al. 2005; see a recent review in Ray et al. 2008). 
Since that time three periods of activity have been observed. 
During the first (the discovery one) the source demonstrated 
five bursts with duration about ten minutes and flux about 1 Jy. 
Bursts followed each other with a period about 77 minutes. 
During the second period just one burst was detected (Hyman et al. 2006). 
The event was similar to those detected before. 
Finally, during the last known period of activity again 
just one burst was detected (Hyman et al. 2007). 
It was weaker ($\sim 0.05$~Jy) and shorter (about 2 minutes) than previous ones.
No counterparts have been found in any other range. 

Several models have been proposed (see the list and references in Hyman et al. 2007): 
a brown dwarf or a cool star, a binary neutron star (NS), a white dwarf, 
a nulling pulsar, a precessing pulsar.
Here I discuss several other possibilities related to not that popular
phenomena which happen 
when a NS passes through some elusive evolutionary stages. Some
of them turn out not to be very likely to explain properties of GCRT
J1745-3009, others are more promising. I focus on  situations when magnetic
field lines of a NS become opened for a short period of time, which
corresponds to a burst, with possible repetition with the 77-min period, or
on situations when an observer can periodically observe opened field lines.

\section{A highly magnetized NS at the propeller or georotator stage}

 During its evolution a NS normally passes through several evolutionary stages 
(see, for example, Lipunov 1992): Ejector, Propeller, Accretor, Georotator.
Radio pulsar activity corresponds to the early phases of the Ejector stage. 
At which stage a NS appears depends 
on relation between several critical radii:
the gravitational capture radius  -- 
$R_\mathrm{G}=2GM/v^2$, 
light cylinder radius -- $R_\mathrm{l}=c/\omega$,
Alfven radius -- $R_\mathrm{A}$,  corotation radius -- 
$R_\mathrm{co}=(GM/\omega^2)^{1/3}$, and the so-called
Shvartsman radius -- $R_\mathrm{Sh}$ (for all details see,
for example, Lipunov 1992).

In some cases one can expect a more tricky situation due to a high
level of asymmetry of a system, or due to rapid changes of parameters of
surrounding medium.
Let us consider a highly magnetized NS (HB-NS)
with $B\sim 10^{15}$~G and velocity 200 km~s$^{-1}$. 
Let us assume that its spin period is 77 minutes 
(such long periods are possible for HB-NSs due to a rapid spin-down
related to a fall-back disc around a NS, see de Luca et al. 2006).
In normal interstellar medium (ISM) 
such an object would be on the 
Propeller stage (Shvartsman 1970, Illarionov \& Sunyaev 1975), 
but on an unusual one, 
as $R_\mathrm{G}<R_\mathrm{A}$. 
So, it is an analogue of the Georotator stage 
(Lipunov (1992) in Ch.6 calls it "a non-gravitating propeller"), 
but with the reversed relation between $R_\mathrm{co}$ and $R_\mathrm{A}$.
  
Here we are interested in the case $R_\mathrm{G}, R_\mathrm{co}<R_\mathrm{A}<R_\mathrm{l}$.
At the magnetospheric boundary gravity is not important, as on the Georotator stage
(Rutledge 2001 calls it the MAGAC stage, Toropina et al. 2001 -- magnetic plow). 
Accretion is impossible as $R_\mathrm{A}>R_\mathrm{co}$. 
Several possibilities interesting in the context of transient radio emission
observation can exist.

{\it Variant 1a.} For 
some combinations of spin, velocity and magnetic dipole vectors, 
due to the ram pressure of the ISM
the magnetosphere can obtain a long ``tail'' similar to the 
magnetotail of the Earth. 
This tail can go beyond the light cylinder.
Obviously, rotation is not important for the tail formation if spin and velocity
are perfectly aligned.
For perfect alignment the ``tail'' does not go beyond the light cylinder
as it is parallel to the spin axis.  

{\it Variant 1b}. The back (respect to the flow) part of the magnetosphere 
is not only elongated, but is also widened. 
For low density of the ISM the magnetospheric boundary 
can approach the light cylinder. 
Again, relative orientations of the three vectors are important.
For example, for the perfect alignment one can apply the analytic solution developed for 
the Earth magnetosphere (Zhigulev, Romishevsky 1959, 
see also Lipunov 1992, Ch. 4).
In this case the size of the magnetosphere in the direction
perpendicular to spin and velocity approaches $2\cdot R_\mathrm{A}$ on the back side.
So, if the density of the surrounding medium is low 
(which is not unexpected due to
formation of a kind of an ``atmosphere'' around 
the magnetosphere due to propeller
action), then the magnetospheric boundary can reach the light cylinder.

Effectively, in both variants there is
 a region of opened field lines  which can result in
detectable radio emission. It is easy to obtain periodicity with the spin period. 
Changes in the ISM properties, 
sporadic accretion from a fall-back disc or an asteroid belt can result in
transient behaviour.

As the magnetospheric boudary approaches the light cylinder, the velocity of its
rotation is close to the velocity of light. 
This is called a ``relativistic propeller'' 
(Lipunov 1992, Ch. 6). 
Such relativistic non-gravitating propeller can be a possible
explanation for the GCRT.  

{\it Variant 1c}.
Romanova et al. (2001) and Toropina et al. (2001) 
presented numerical studies of the Georotator stage. 
At this stage they observe reconnection in 
the magnetotail a NS. For the non-rotating case they provide an estimate of the
total energy: 

\begin{equation}
E= 10^{28} B_{15} n^{1/2} v_{200} \, {\rm erg}.
\end{equation}

For two limiting cases duration of an event is $t_\mathrm{rec}=10^5 B_{15}^{1/3}n^{-1/6}v_{200}^{-4/3}\, {\rm sec}$
for high matter density, and $t_\mathrm{rec}=74 B_{15}^{1/3}n^{-1/6}v_{200}^{-1/3}\, {\rm sec}$ -- for low density.
Few minutes periodicity 
observed in the case of GCRT is well within this interval. However,
energy is too low to explain an event at
further than few tens of parsec, even of all energy
goes into radio emission at $\nu \sim 330$~MHz.


 \section{Transient propeller}

Several authors discussed the following possibility. If on the Propeller
stage cooling in the matter around the rotating magnetosphere is efficient
enough (for example, synchrotron emission of thermal electrons on the
magnetic field frozen in plasma can be important, the main problem here is
that the frequency of the synchrotron emission can be smaller than the
plasma frequency for large magnetospheric radius), than not a static
atmosphere, but a dense envelope with growing mass is formed on top of the
magnetospheric boundary (see the references and discussion in Lipunov 1992).

The radius of the boundary starts to be determined by 
the balance between the magnetic field  and 
the weight of the envelope:
$
\mu^2/8\pi R^6 = GM_\mathrm{sh}M/4 \pi R^4.
$

As the mass of the shell, $M_\mathrm{sh}$, 
grows the radius of the magnetosphere decreases 
till it reaches the corotation radius. 
Then the whole envelope collapses to the NS. 
At this moment an X-ray and a radio bursts can be expected.
The duration of a burst can be estimated as:
$\Delta t_\mathrm{b}=R_\mathrm{co}/v_\mathrm{ff}= P/2\sqrt{2} \pi,$
$v_\mathrm{ff}$ -- the free-fall velocity.

{\it Variant 2a}. An isolated NS.
In this case the 77-minute period should be associated with 
the interval between two successive episodes of an envelope collapse.
This time can be estimated as:
\begin{equation}
\Delta t=M_\mathrm{sh}/\dot M = 
10^8 \, \mu_{30}^2 P^{-4/3} \rho_{-24}^{-1} v_{100}^3 \, {\rm s}.
\end{equation}

A 77 minute spin period would lead for some parameters to 
77 minute interval between successive collapses. Then, then crossing time
(the free-fall time) would be about $2 \pi$ times smaller, as it is roughly
observed for GCRT J1745-3009.

{\it Variant 2b}. A NS in a binary.
Of course, the regime with a collapsing envelope can be reached also in a binary system. 
There the accretion rate can be much higher than for an isolated NS, 
so the 77-minute periodicity can be reached for shorter periods 
(or, for higher magnetic fields).

 \section{Ejector in a binary system}

In a binary system when a NS is on the Ejector stage 
and the companion is a normal (hydrogen burning) star 
several situations can appear in which a transient radio source can appear.

{\it Variant 3a}. Superejector.

 It can happen that a companion starts to fill its Roche lobe 
when a NS is still very rapidly rotating. 
In this case the so-called Superejector stage can be set on.
``Super'' stands for superEddington accretion rate (Lipunov 1992)
(here, the accretion rate is used just in terms of properties 
of the matter flow from the normal star; 
of course, no accretion on the surface of a NS is going on). 
 This stage is possible for very rapid rotators:
$
P_\mathrm{SE}<11~{\rm msec} \, \mu_{30}^{4/9}. 
$

On this stage a NS is surrounded by a cavern produced by 
the radio pulsar emission. 
Due to a sudden decrease of the accretion rate the cavern can open, 
so a radio burst can be observed. 
The stage is not well studied, so it is difficult to make predictions 
about periodicity in such outbursts.

{\it Variant 3b}. Pulsating cavern.
 The case of an Ejector in a binary for
 non-critical accretion rates
was studied by Lipunov and Prokhorov (1984). 

At this stage $R_\mathrm{Sh}>R_\mathrm{G}$. No dense envelope is formed, 
and the cavern boundary is define by the stellar wind pressure 
and pulsar wind (and radiation) pressure.

 Let us estimate the characteristic time scale for cavern pulsations
following the general line presented by Lipunov and Prokhorov (1984).
One can  start with an equation of the pressure equilibrium:

\begin{equation}
\frac{\dot M V_\mathrm{w}}{4 \pi R_0^2} \cos^2\psi + P_\mathrm{g}=
\frac{L_\mathrm{m}}{4 \pi R^2 c}\cos ^2 \chi + 
\frac{\delta L_\mathrm{m}t}{3V}.
\end{equation}
The first term on the left side represents the ram pressure of the stellar
wind. 
Here $\dot M=\alpha L_0/(V_\mathrm{w}c) $ -- is mass loss by the normal star.
$V_\mathrm{w}$ -- stellar wind velocity. $ P_\mathrm{g}$ -- is gas pressure,
in an isothermal flow one can use  $P_\mathrm{g} = A R_0^{-2}$. $R_0$ and
$R$ are distances from the normal star and a NS to a point on the cavern,
respectively. $\psi$ and $\chi$ are angles between the normal to the cavern
surface and vectors $\vec R_0$ and $\vec R$, respectively. 
On the right side the first term shows the pressure due to the relativistic
wind from a NS. $L_\mathrm{m}$ -- is the NS luminosity, which can be estimated,
for example, with the magneto-dipole formula. 
If the cavern is closed then $\delta=1$, otherwise it is zero. 
The term with $\delta$ 
appears as radiation and particles are collected inside the cavern
during the time interval when the cavern is closed.
The volume, $V$, can be roughly estimated as $(4/3)\pi  R_-^3$, where $R_-$
is the distance towards the back end of the cavern on the line connecting the
star and the NS.

It is necessary to estimate the critical time, $t_\mathrm{br}$
 when the cavern becomes opened.
At the critical point corresponding to $R_-$ the first term on the left side
is zero.  For the gas pressure I use $A=k \dot M T_* /(2 \pi V_\mathrm{w}
m_\mathrm{p})$, $R_0=a+R_-$, $a$ -- is the semi-major axis of the binary
system, $k$ -- the Boltzmann constant.

 The critical radius of the cavern can be estimated from 
$P_\mathrm{g}=L_\mathrm{m}/(4 \pi R_-^2 c)$ when $\delta=0$.
Then one can estimate $t_\mathrm{br}$ from $P_\mathrm{g}=L_\mathrm{m}t_\mathrm{br}/3V$
(in this case it is assumed that two terms on the right side of eq. 3 are
of the same order):
\begin{equation}
t_\mathrm{br}=(R_\mathrm{Sh}/c) \, (1/(\sqrt{k_*}-(R_\mathrm{Sh} /a))), 
\end{equation}
here 
$k_*=4 \pi A/(\dot M V_\mathrm{w})$. 
This time interval can be equal to 77 minutes for some parameters of a NS
and a binary system.

{\it Variant 3c}. Floating cavern.
 This case corresponds to $R_\mathrm{Sh}<R_\mathrm{G}$. A dense envelope
is formed, which defines the boundary.
Lipunov and Prokhorov (1984) discussed a possibility that 
a cavern around a NS in a binary can become detached from the system, 
and ``sail'' away carried by the stellar wind.  This happens when an expanding cavern 
reaches the $R_\mathrm{G}$.

 In this case (Lipunov, Prokhorov 1984) the duration of the burst is:

\begin{equation}
\Delta t_\mathrm{b}= R_\mathrm{G}/c\approx 9\, (v/50\, {\rm km/s})^{-2}\, {\rm min}.
\end{equation} 
Clearly, only a very slow wind can produce a long burst.

The interval between bursts is: $
\Delta t= (a/R_\mathrm{Sh})^2 R_\mathrm{G}/c.
$
Here $a$ is slightly larger than $R_\mathrm{Sh}$, so it is easy to have $\Delta t$ few times
larger than $\Delta t_\mathrm{b}$.


\section{Populational aspects}

 If the GCRT is a typical representative of its class, then one can conclude
that there are no more sources of this type in the direction of the search
as the source was detected significantly above the threshold and
observations are long in comparison with the duty cycle of the source. 

 If the source is close, then one can expect to find more weaker bursts from
other slightly further objects, unless we are extremely lucky to have a rare
source close to us. So, the GCRT is most probably located close to the
center of our Galaxy. Basing on the coverage of the galactic plane and the
bulge by the observations which resulted in the detection, one can expect to
have
$\sim$~100-1000 such sources in the Galaxy (for isotropic emission, if
emission is beamed, then the number of sources can be higher).

The GCRT can be an isolated NS.
There are $\sim 10^9$~NSs in the Galaxy, their birth rate is $\sim
1/50$~yr$^{-1}$. If the GCRT stage is something typical to most of NSs, then
the duration of the stage should be about $10^3$ years. This is very short
time scale for any stage of a normal NS evolution. So, one needs to proceed
assuming that just a fraction of NSs pass through this stage of activity.

HB-NSs are assumed to represent few percents of the total
population, may be slightly more. 
If a typical HB-NS at some stage of its evolution demonstrate an
GCRT-type activity, then the duration of such stage should be about
$10^5$~yrs. HB-NSs are known to evolve rapidly, i.e. their spin-down scale
is very short in their young years. They are expected to reach relatively
quickly the ultimate stage of their evolution (Accretor or Georotator). When
they reach such a stage it
is difficult to associate such short timescale as
$10^5$~yrs with such objects. An intermediate stage for normal field NSs 
-- the Propeller stage or its
modifications -- are, however, promising from the populational point of view,
as for this stage the timescale about $10^5$~yrs is typical.

\section*{Acknowledgments} 
The work was supported by the RFBR grant 07-02-00961.
The author thanks Dr. M.E. Prokhorov for discussions and the Organizers for
hospitality.



\begin{thebibliography}{99}
\bibitem{d2006}
de Luca A., et al., 2006, Science, 313, 814
\bibitem{h2005}
Hyman S.D., Lazio T.J.W., Kassim N.E., Ray P.S., Markwardt C.B., Yusef'Zadeh F., 2005, 
Nature, 434, 50
\bibitem{h2006}
Hyman S.D., Lazio T.J.W., Roy S., Ray P.S., Kassim N.E., 2006, ApJ, 639, 348 
\bibitem{h2007}
Hyman S.D. et al., 2007, ApJ, 660, L121
\bibitem{il1975}
Illarionov A.F., Sunyaev R.A., 1975, A\&A, 39, 185
\bibitem{l1992}
Lipunov V.M., 1992, ``Astrophysics of neutron stars'', Springer-Verlag, Berlin
\bibitem{lp1984}
Lipunov V.M., Prokhorov M.E., 1984, Ap\&SS, 98, 221
\bibitem{ray2008}
Ray P.S., et al., 2008, arXiv: 0808.1899
\bibitem{rttl2001}
Romanova M.M., Toropina O.D., Toropin Yu.M., Lovelace R.V.E., 2001,
Proc. of the 20th Texas Symp. on Relativistic Atrophysics, Eds. K. Wheeler, 
H. Mortel, AIP conf. proc. vol. 586, p. 519
\bibitem{r2001}
Rutledge B.E., 2001, ApJ, 553, 796
\bibitem{s1970}
Shvartsman V.F., 1970, Radiofizika, 13, 1852
\bibitem{t2001}
Toropina O.D., Romanova M.M., Toropin Yu. M., Lovelace R.V.E., 2001, ApJ, 561, 964
\bibitem{zr1959}
Zhigulev V.N., Romishevskii E.A., 1959, Doklady Akad. Nauk, 127, 1001 
\end{thebibliography}
\end{document}